# A LOW-CHARGE DEMONSTRATION OF ELECTRON PULSE COMPRESSION FOR THE CLIC RF POWER SOURCE

R. Corsini, A. Ferrari, J.P. Potier, L. Rinolfi, T. Risselada, P. Royer, CERN, Geneva, Switzerland

*Abstract*

The CLIC (Compact Linear Collider) RF power source is based on a new scheme of electron pulse compression and bunch frequency multiplication using injection by transverse RF deflectors into an isochronous ring. In this paper, we describe the modifications needed in the present LEP Pre-Injector (LPI) complex at CERN in order to perform a low-charge test of the scheme. The design of the injector (including the new thermionic gun), of the modified linac, of the matched injection line, and of the isochronous ring lattice, are presented. The results of preliminary isochronicity measurements made on the present installation are also discussed.

## 1 INTRODUCTION

The time structure of the CLIC drive beam is obtained by the combination of electron bunch trains in rings using RF deflectors [1]. The next CLIC Test Facility (CTF3) at CERN will be built in order to demonstrate the scheme and to provide a 30 GHz RF source with the nominal parameters [2]. CTF3 will be installed in the area of the present LPI complex. As a preliminary stage, the existing installation will be modified in order to perform a test of the combination scheme at low charge. The layout of the LPI complex after the modifications is shown in Figure 1. The first part of the LEP Injector Linac (LIL) will be dismantled and shielding blocks will be added, creating an independent area that can be used for component tests and later for the commissioning of the new CTF3 injector. The LIL bunching system will be moved downstream and a new gun [3] will be installed. Eight of the sixteen accelerating structures of LIL, the extraction lines to the PS complex, and the positron injection line between LIL and EPA will be removed.

## 2 THE INJECTOR AND THE LINAC

The new triode gun has a design voltage of 90 kV. Its control grid can be pulsed in order to provide a train of up to seven pulses with a repetition rate of 50 Hz. The nominal pulse length is 6.6 ns FWHM, and the pulses are spaced by 420 ns, corresponding to the Electron Positron Accumulator (EPA) circumference. The nominal peak current of the gun is 1 A. The present bunching system of LIL, composed of a single-cell pre-buncher and a standing wave buncher, will be used. It will be powered by a 30 MW klystron and will provide a 3 GHz bunched beam at 4 MeV, with a normalised rms emittance of 50 $\pi$ mm mrad. Each pulse will then be composed of 20 bunches each with a charge of 0.1 nC and a length of about 10 ps FWHM. These values are extrapolated from measurements made on the present installation [4].

The pulse train will be accelerated to a maximum energy of 380 MeV, using eight travelling wave accelerating structures, powered in groups of four by two 40 MW klystrons.

The beam parameters have been chosen to minimise the energy spread generated by beam-loading in the LIL structures, while still keeping a charge per bunch which is high enough to give a good resolution for the streak camera measurements of the beam time structure. The beam-loading parameter in LIL is 0.2 MeV/nC per structure. The resulting energy spread within each pulse is about 3 MeV. An additional energy difference of roughly 3 MeV will occur between the first two pulses, i.e., before the steady state is reached.

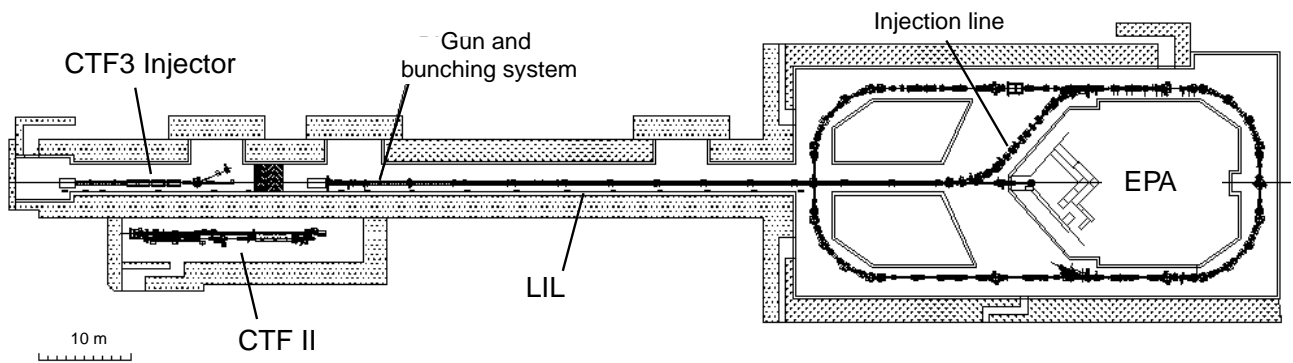

Figure 1: Layout of the LPI complex after the modifications planned for the CTF3 preliminary phase.

The total energy spread of about 6 MeV is within the EPA acceptance (± 1% total). It can nevertheless be reduced by a factor two by delayed RF filling of the structure or by dumping two out of the seven pulses of the train. For the frequency multiplication test, one needs five pulses maximum.

The linac optics will be adapted to the new layout. Two quadrupole triplets, one located after the bunching system and the other between the first two acceleration structures, will provide matching to the modified LIL FODO lattice. The last two LIL structures will be removed and replaced by a matching section to the injection line. The design of the linac optics and of the matching sections is presently in progress. Apart from the new gun, no new equipment is needed and only a re-arrangement of the existing components suffices.

## 3 THE INJECTION LINE

The present line from LIL to EPA is achromatic in both planes. In addition to the main horizontal bending magnets, the line contains two small vertical dipoles, since the levels of LIL and EPA differ by 15 cm in order to allow injection from the inside of the ring. However, the line must also be made isochronous, in order to preserve the bunch length from the linac to the ring. This is essential for the combination process, for which short bunches (< 20 ps FWHM) are needed. Furthermore, the line has to be re-matched to the new ring lattice. A new optics of the line has been found which satisfies the requirements of the CTF3 preliminary phase.

If R is the 6x6 dimensional transfer matrix of the injection line, the achromatic condition implies that the matrix elements $R_{16}$, $R_{26}$, $R_{36}$ and $R_{46}$ are equal to zero. To preserve the bunch length, the elements $R_{51}$, $R_{52}$, $R_{53}$, $R_{54}$ and $R_{56}$ must be small. In particular this last condition is not satisfied in the present line, for which $R_{56}$ = -1.05 m (against a requirement of $|R_{56}|$ < 0.1 m). $R_{56}$ will be controlled in the new arrangement by three high-gradient quadrupoles, placed between the bending magnets and the septa. The eight other coefficients, as well as the two transverse lattice functions $\beta_x$ and $\beta_y$, can be controlled by five quadrupoles in the straight section of the injection line (the gradients and the positions of these five quadrupoles were chosen independently in order to fulfil all the requirements).

The dispersion and the $\beta$-functions of the new injection line are displayed in Figure 2. The Twiss parameters at the entrance of the line are the ones required for the transverse matching at the injection point in the ring ($\beta_x$ = 31 m, $\alpha_x$ = -2, $\beta_Y$ = 5 m, $\alpha_Y$ = -1). The beam envelope is within the acceptance of the line. The $R_{51}$, $R_{52}$, $R_{53}$, $R_{54}$ and $R_{56}$ coefficients remain small enough for no bunch lengthening to occur.

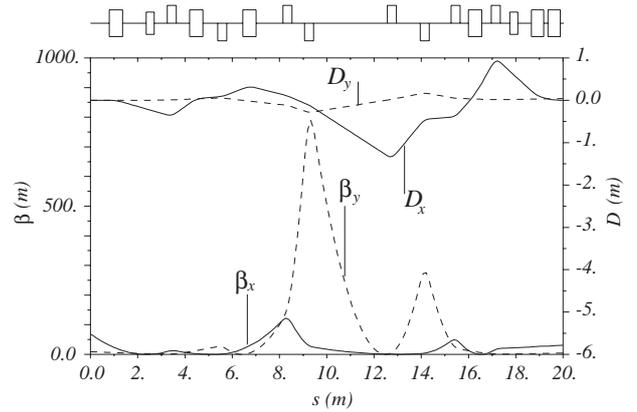

Figure 2: Optics functions of the new injection line.

Two quadrupoles must be added to the present layout, and some of the existing quadrupoles must be moved. All quadrupoles in the line will be fed with independent power supplies. However, the geometry of the line is preserved, such that no major hardware modifications are needed. The quadrupoles and power supplies from the dismantled beam lines can be reused in the new injection line.

## 4 THE RING LATTICE

The lattice of the EPA ring will be modified to become isochronous and thus to preserve the bunch length and spacing during the combination process (three to five turns).

A first isochronicity test has been performed in the present ring [5]. The isochronous lattice was obtained by changing the strength of the quadrupoles without making any hardware modification. Measurements of the beam time structure were made using a streak camera. The bunch length increased rapidly over a few turns in the normal case, while no significant bunch lengthening was observed over 50 turns in the isochronous case. An evaluation of the momentum compaction has been obtained by measuring the bunch spacing, yielding values of $\alpha$ as small as $2.3 \times 10^{-4}$, close to the goal of the future CTF3 ($|\alpha| \leq \pm 10^{-4}$).

The dispersion in the modified isochronous lattice is shown in Figure 3, together with that of the normal lattice used for LEP operation. In the isochronous case, the dispersion varies strongly in the straight sections. The injection line was therefore badly matched to the ring, and the larger beam envelopes gave rise to beam losses on the first few turns. The situation will be worse when the RF deflectors are installed, since they will reduce the available aperture.

To solve this problem, in the new isochronous lattice the dispersion is made zero in the straight sections [6]. The dispersion and the $\beta$-functions are shown in Figure 4. This needs the displacement of four quadrupoles, and the decoupling of one of the existing families.

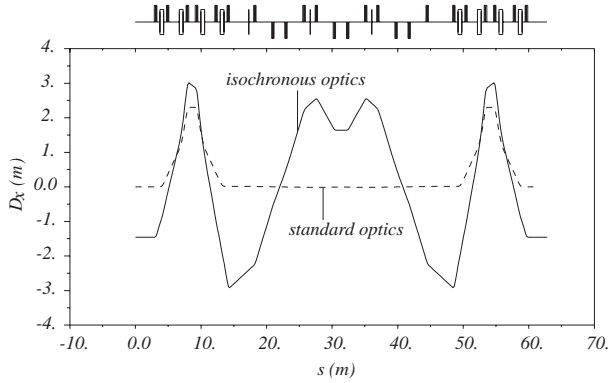

Figure 3: Dispersion in the present EPA ring, for different optics (half ring shown).

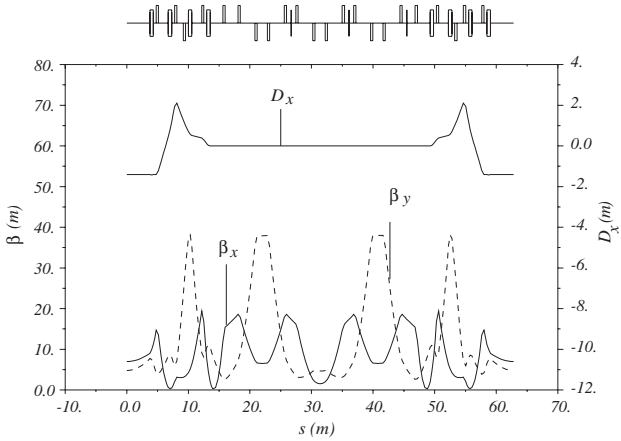

Figure 4: Optics functions of the new EPA ring lattice (half ring shown).

## 5 THE COMBINATION TEST

Two transverse RF deflectors will replace the present fast injection kickers. They will create a time-dependent closed bump of the reference orbit, allowing interleaving of three to five bunch trains (see Figure 5). The combination test requires $C = n\,(\lambda \pm \lambda/N)$ where n is an integer, $C$ is the ring circumference, $N$ is the combination factor and $\lambda$ is the RF wavelength in the deflectors and the linac. Combination factors of 3, 4 and 5 will be tested in the preliminary phase of CTF3. One arc of the ring will be displaced by 7.5 mm, in order to fulfil the condition with $N = 4$. The other combination factors can be tested by changing the RF frequency by $\pm$ 150 kHz. The bandwidth of the klystrons is wide enough to cover this range. The accelerating structures will be tuned in operation to follow the change of the RF frequency, by varying their temperature $\pm$ 3 °C. The RF deflectors are travelling wave iris-loaded structures, for which the resonant mode is a deflecting mode with a $\pi/2$ phase advance per cell and a negative group velocity. They have been used in the past to measure the bunch length in LIL [7]. Each one is 27 cm long, with 6 cells and an iris diameter of 2.3 cm.

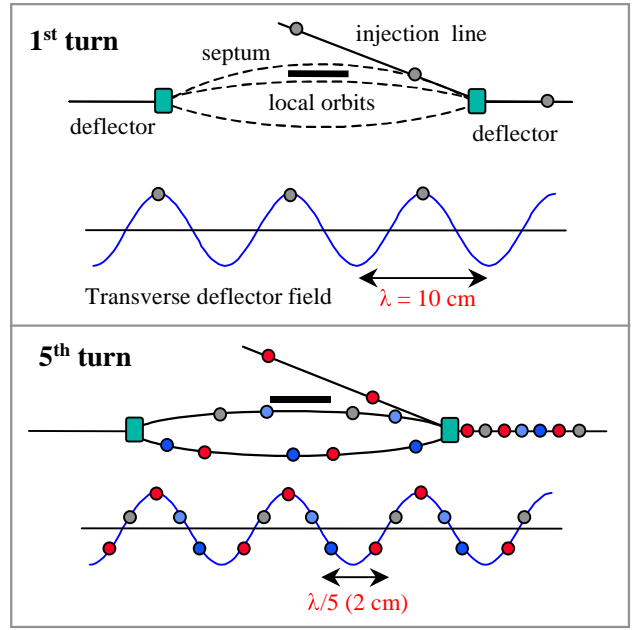

Fig. 5: Principle of injection by transverse RF deflectors, for a combination factor of 5. After injection, the circulating bunches follow orbits that are inside the septum. After five turns the beam is extracted.

Each will be powered by one of the existing 30 MW klystrons and a power of about 4 MW each is needed for the nominal deflecting angle of 2 mrad at 380 MeV. While one of the fast injection kickers will be removed, the other will be displaced but kept in the ring to allow conventional single-turn injection. It will be used during commissioning, to check the ring optics prior to the installation of the RF deflectors. Also, one of the positron injection kickers, located at the opposite side of the ring, will remain in place and will be used to extract the circulating beam to a dump.